\documentclass[aps,pre, twocolumn, groupedaddress]{revtex4-1}

\usepackage{graphicx}
\usepackage{dcolumn}
\usepackage{amsmath}    
\usepackage{amssymb}
\usepackage{bm} 
\usepackage{hyperref}
\usepackage{latexsym}
\usepackage{verbatim}
\usepackage{color}

\def\beq{\begin{equation}}
\def\eeq{\end{equation}}

\def\etal{{\it et al.}}

\def\beq{\begin{equation}}                           
\def\eeq{\end{equation}}                           
\def\bea{\begin{eqnarray}}                           
\def\eea{\end{eqnarray}}        

\preprint{}

\begin{document}

\preprint{}


\title{Density phase separation and order-disorder transition in a collection of polar self-propelled particles}
\author{Sudipta Pattanayak}
\email{pattanayak.sudipta@gmail.com}
\affiliation{S N Bose National Centre for Basic Sciences, J D Block, Sector III, Salt Lake City, Kolkata 700106}
\author{Shradha Mishra}
\email[]{smishra.phy@itbhu.ac.in}
\affiliation{Department of Physics, Indian Institute of Technology (BHU), Varanasi, India 221005}

\date{\today}

\begin{abstract}
{We study the order-disorder transition in a collection
of polar self-propelled particles, interacting through a distance
dependent short range alignment interaction. 
A distance dependent  interaction parameter $a_0$ is introduced such that 
on decreasing $a_0$  interaction decay faster with distance $d$ and  for $a_0=1.0$ model
reduces to Vicsek's type. 
For
all $a_0>0.0$, system shows a transition from disorder to long ranged ordered 
state.  We find  another phase transition 
from phase separated to nonphase separated state with decreasing $a_0$: at the same time
order-disorder transition changes from discontinuous to continuous type. Hence density phase 
separation plays an important role in predicting the nature of order-disorder transition.
We also calculate the two-point density structure factor 
using coarse-grained hydrodynamic equations of motion  
with an introduction of a density dependent alignment term in the equation introduced by
Toner and Tu \cite{tonertu}. Density structure factor shows a
divergence at a critical wave-vector $q_c$, which  decreases with decreasing 
density dependent alignment term.
Alignment term in the coarse-grained equation plays 
 the same role as the distance dependent parameter 
$a_0$ in the microscopic simulation.
Our results
can be tested  in many biological systems: where
particle have tendency to interact strongly with their closest neighbours.}
\end{abstract}
\maketitle
\section{Introduction \label{introduction}}
 Flocking \cite{bacterialcolonies, insectswarms, birdflocks, fishschools}
 - the collective, coherent motion of large
number of organisms, is one of the most familiar and ubiquitous biological phenomena.
Last one decade there have been an increasing interest in  rich behaviour of
these systems which are far from equilibrium \cite{sriramrev3, sriramrev2, sriramrev1}.
One of the key feature of these flocks, is that the systems show a transition from 
disordered state to a long ranged ordered state with the 
variation of system parameters: like density, noise strength etc. \cite{benjacob, chate2007, chate2008}.
Nature of such transition is a matter of debate even after many years of introduction
of a minimal model  by T. Vicsek {\em et al.} 
in 1995 \cite{vicsek1995}, also called as Vicsek's model (VM). 
 In the model  a collection of point particles
 move along their heading direction and  
align with  
 their neighbours lie in a small metric distance.
Many studies are  done with other  model called
as topological distance model, where  particles
interact through topological distance \cite{chatetopo, chatetopo1}.
Initially Vicsek's 
study on metric distance model finds the transition is continuous \cite{vicsek1995} but later studies of 
\cite{chate2007, chate2008} find it discontinuous.  
Similary for topological distance model, study of \cite{biplab} claims
dicontinuous  but  in \cite{chatetopo} finds the transition is continuous.
Hence nature of transition is a matter of curiosity in polar flock. 

In our present study we ask  a {\it question}, what causes the nature
of transition to change from discontinuous type to continuous one ?
In previous studies of  metric as well as topological distance models,
particles interact with the same interaction strength within an interaction
metric or topological distance. But in many biological systems
particles have tendency to interact more with their closest neighbours.
In recent study of \cite{andrea} using maximum entropy principle, they find the  functional dependence of the interaction
on the distance which decays exponentially over a range of few individuals.

In our model we  introduce a distance dependent interaction parameter $a_0$, 
such that interaction decays with distance
within a small metric distance. 
For $a_0=1.0$, interaction is of Vicsek's type and as we decrease
$a_0$ strength of interaction decays faster with distance.
For all  non-zero interaction parameter $a_0 >0.0$
system shows a disordered state at small density, high noise
strength and long-ranged ordered state at high density low noise 
strength. We also find  another phase transition from 
phase separated to nonphase separated  
state as we decrease $a_0$. Order-disorder
transition is first order for phase separated state and gradually becomes 
continuous as we approach nonphase separated state.\\
 
In rest of the article,  in section \ref{Model} we  introduce the microscopic
rule based model for distance dependent interaction in the Vicsek's model 
and then  write the phenomenological hydrodynamic equations of motion
for a  collection of polar self-propelled particles. 
 Numerical details of microscopic   simulation are givem in section \ref{Numerical study}.
Section \ref{results} gives the results of numerical simulation and linearised calculation.  
Finally in section \ref{discussion} we discuss our main results and
future prospect of our study. Detail calculation of linearised structure factor
is given at the end in the appendix \ref{App:AppendixA}.

\section{Model \label{Model}}
We study a collection of polar self-propelled particles on a two-dimensional substrate. 
These particles interact through a short range alignment interaction
which decays with distance inside a small interaction radius. We first
describe a rule based distance dependent model for such system, which is similar
to model introduced by Vicsek's but with an additional distance dependent interaction. And then  we write coupled hydrodynamic equations of motion for density and velocity derived 
from the microscopic model.\\
{\it Microscopic Model}:
Each particle in the collection is defined by its position ${\bf r}_i(t)$ and
orientation $\theta_i(t)$ or unit direction vector ${\bf n}_i(t)=[\cos\theta_i(t), \sin\theta_i(t)]$
on a two-dimensional substrate. Dynamics of particle is given by two updates. One for the position, that
takes care  of its self-propulsion and other for orientation, that cares about the interaction between particles.
Self-propulsion, is introduced as a motion towards its orientation direction with some fixed
step size. 
Hence position update of particles,
\begin{equation}
{\bf r}_{i}(t+1)={\bf r}_{i}(t)+v_{0}{\bf n_{i}}
\label{eqn1}
\end{equation} \\
and orientation update with a distance dependent short range alignment interaction
 \begin{equation}
{\bf n_{i}} (t+1)=\frac{\sum _{j\in R_{0}} {\bf n_{j}}(t)a_{0}^{d_{ij}}+N_{i}(t)\eta {\bf \zeta}_{i}}{W_{i}(t)}
\label{eqn2}
\end{equation} \\
 where sum is over all particles inside the interaction radius with
$\vert {\bf r}_{j}(t)-{\bf r}_{i}(t)\vert <1$,
 $N_i(t)$ is number of particle within unit interaction radius and $W_{i}(t)$ 
is the normalisation factor, which makes ${\bf n}_i(t+1)$ again
a  unit vector, $\eta$ is the strength of noise, which
we vary between  zero to $1$ and ${\bf \zeta}_i(t)$ is a random unit vector. \\
{\it Phenomenological hydrodynamic equations of motion :}
We also  write the phenomenological hydrodynamic equations
of motion which are  either derived from the above rule based model or written
by symmetry of the system. Density: because total number of particles are
conserved and velocity: is a broken symmetry variable in the ordered state, 
are two hydrodynamic variables in our system. They are defined by
\begin{equation}
\rho({\bf r},t)=\sum_{i=1}^{N}\delta({\bf r}-{\bf r}_{i})
\label{eqn3}
\end{equation}
and 
\begin{equation}
{\bf V}({\bf r},t)=\dfrac{\sum_{i=1}^{N}{\bf n}_{i}(t)\delta({\bf r}-{\bf r}_{i})}{\rho({\bf r},t)}
\label{eqn4}
\end{equation}
Coupled hydrodynamic  equation of motion for density is
\begin{equation}
\partial_{t}\rho=v_{0}{\nabla}.(\rho {\bf V} )
\label{eqn5}
\end{equation}
and for  velocity
\begin{equation}
\begin{aligned}
\partial _{t} {{\bf V}}={} & \alpha (\rho){\bf V} -\beta (\mid V \mid)^2 {\bf V} - \frac{v _{1}}{2\rho _{0}} {\nabla} \rho  \\ & + D_{p} \nabla^{2}{\bf V} -\lambda_{1} ({\bf V}.{\nabla}) {\bf V}-\lambda_{2}({\bf \nabla}. {\bf V}){\bf V} \\ & -\lambda_{3} \nabla (\mid V\mid ^{2})+ {\bf f _{V}}
\end{aligned}
\label{eqn6}
\end{equation} 
These equations are similar to the equations introduced by Toner and Tu for polar self-propelled
 flocks \cite{tonertu}.
Density equation Eq.\ref{eqn5} is a continuity equation, where $v_0$ is the 
self-propulsion speed of the particles. First two terms in the 
velocity equation Eq.\ref{eqn6} is  a mean-field order disorder term.
In general when derived from any metric distance model like Vicsek's model
both $\alpha(\rho)$ and $\beta$ are functions of density: such that 
$\alpha(\rho)$ changes sign at some critical density $\rho_c$. Hence
 homogeneous equations has a disordered state  $V_0=0$ for $\rho_0 < \rho_c$ and 
ordered state $V_{0} = \sqrt{\frac{\alpha(\rho_0)}{\beta}}$ for $\rho_0 > \rho_c$, where $\rho_0$
is the mean density of the system. Distance dependent alignment interaction, which is in
general non-linear,  introduces non-linear density dependence of $\alpha(\rho)$. 
Hence we keep general density dependence of $\alpha(\rho)$.
$v_{1}$, $D_{V}$ and $\lambda$'s are  
 constants, 
$\nabla \rho$ is pressure term and 
$D_{V}$ is the viscosity term.
$\lambda$'s  are convective non-linearities, typically present in fluid 
flow and present here because our velocity field can also flow.
Presence of all three non-linearities show the absence of Galilean invariance 
in polar flock system.
The ${\bf f}_V$ term is a random Gaussian white noise, with zero
mean and variance  
\begin{equation}
<f_{V_{i}}({\bf r},t)f_{V_{j}}({\bf r}^{\prime},t^{\prime})>=2\Delta _{0}\delta_{ij} \delta^{d}({\bf r}-{\bf r}^{\prime})\delta (t-t^{\prime})
\label{eqn7}
\end{equation} \\
where $\Delta_{0}$ is a constant and ($i,j=1,2$) denoting Cartesian components. 

\section{Numerical details \label{Numerical study}} 
\begin{figure}[ht]
\centering
\includegraphics[width=1.0\linewidth]{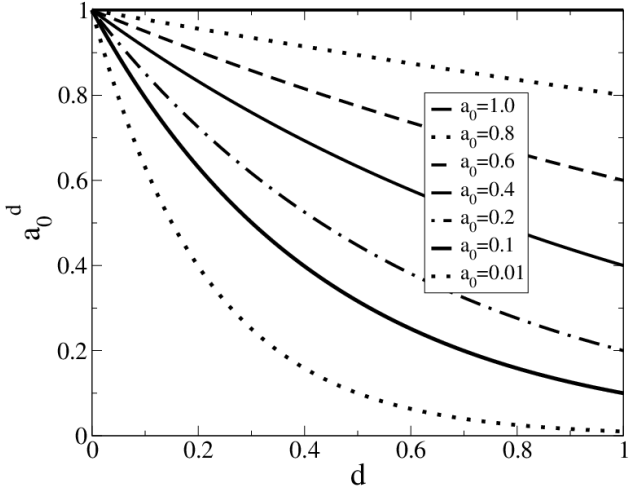}
\caption{Plot of interaction strength ($a_{0}^{d}$) vs. metric distance between particles ($d$) within unit interaction radius for different distance dependent parameter ($a_{0}$=1,0.8,0.6,0.4,0.2,0.1,0.01) in decreasing order from top to bottom. For $a_{0}=1$ all the particles will interact with same interaction strength within unit interaction radius, as we decrease $a_{0}$ effective range of interaction decreases .}
\label{fig:fig1}
\end{figure}

\begin{figure*}[ht]
\centering
\includegraphics[width=1.0\linewidth]{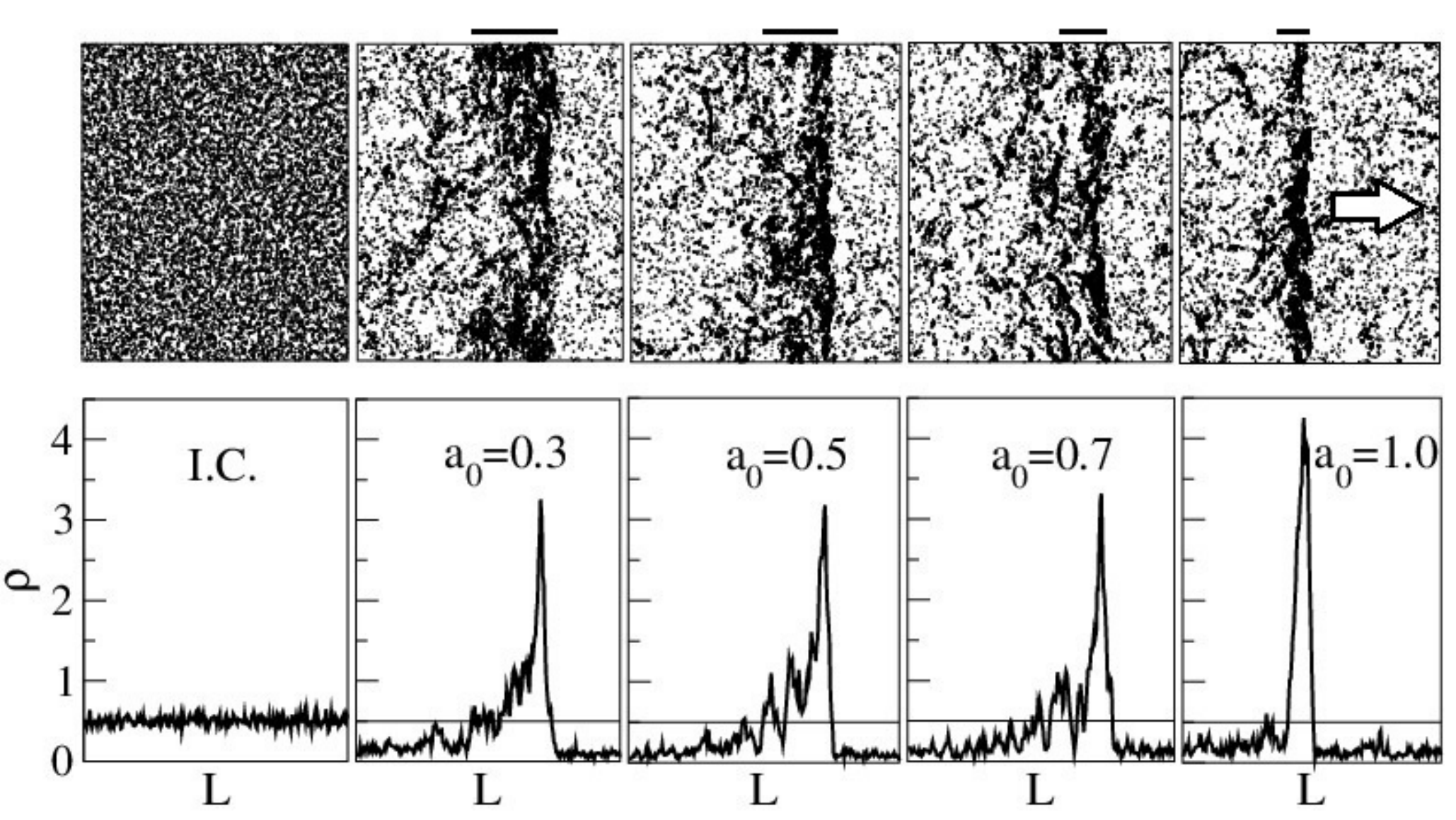}
\caption{Upper panel : We plot real space snap shots of particle distribution. In the left most figure we plot initial condition for all distance dependent parameter ($a_{0}$). Other four plots are the real space particle distribution in steady state for $a_{0}=0.3,0.5,0.7,1.0$ from left to right. We draw a horizontal line in top of each real space snap shot which shows approximate length scale of the band and in right most figure we draw an arrow, which indicates the direction of movement of the band for all $a_{0}$. Lower panel : We plot density distribution ($\rho$) in horizontal direction or density distribution along the direction of band formation. Left most figure shows initial random density distribution for all $a_{0}$. Other four figures show density distribution for $a_{0}=0.3,0.5,0.7,1.0$ from left to right. In each figure we draw an horizontal line at $\rho =0.5$, mean density of the system. Here $\rho$ is the density along band direction and L is the system size. System parameters we have used, $L=256, \rho_0=0.5$ .}
\label{fig:fig2b}
\end{figure*}
We numrically study the  microscopic model introduced in Eqs.\ref{eqn1} and \ref{eqn2} for different distance dependent interaction parameter $a_0$. 
Form of interaction potential
is shown in Fig. \ref{fig:fig1} for different $a_0$ as a function of distance. For $a_0=1.0$,
all the particles within the interaction radius interact with same strength, but as we
decrease $a_0$ effective range of interaction decreases. We vary $a_0$ from $1.0$ to small value
$0.01$ and  for $a_0 =0.0$ (no alignment interaction). We keep the speed $v_0 = 0.5$ of the particles. 
 We start with initially homogeneous
density and random orientation of the particles on a two dimensional lattice of size $L \times L$
 and mean density $\rho_0$ with periodic boundary condition.\\
System shows a phase transition from disordered to long-ranged ordered  state
with the variation of noise strength $\eta$. 
Ordered state is characterised by 
 global velocity defined by 
\begin{equation}
 V=\vert\frac{1}{N} \sum _{i=1} ^N {\bf{n}}_{i}(t)\vert.
\label{eqn27}
\end{equation}
Typical plot of $ V $ vs. $\eta$  is shown in cartoon picture in inset of Fig \ref{fig:fig7}
We study our model in three different regions $I$(disordered), $II$ (close to transition:on the ordered side)
and $III$ (deep ordered state)  of phase diagram.
First we study the model for region $II$ (close to transition). 
Since effective range of interaction as shown in Fig \ref{fig:fig1}
decreases with $a_0$, hence for fixed density, critical value of 
noise strength also changes. For each  $a_0 = 1.0,0.8,0.7,0.6,0.5,0.4,0.3,0.2,0.15,0.1,0.01$ we first estimate
the critical $\eta_c(a_0)$, then we choose value of $\eta(a_0)=\eta_c-\delta \eta(a_0)$, such that
 system have approximately same global velocity in the steady state. List of values of $\eta(a_0)$ used
in region $II$ of phase diagram is 
given in Table \ref{table:table1}. 
We choose noise  strength ($\eta$) in region $I$  and $III$, $0.7$ and $0.1$ respectively such that system
is in the disordered/ordered state for all $a_0$. 

\begin{table*}[ht]
\begin{tabular}{ |p{2.5cm}||p{2.5cm}|p{2.5cm}|p{2.5cm}|p{3cm}|  }
 \hline
 \multicolumn{5}{|c|}{Distance Dependent Interaction} \\
 \hline
Distance dependent parameter($a_{0}$) & Band density($\rho_{b}$) & Band width($W_{b}$) & Fraction of particle forming band ($n_{c}$) & Value of Noise-strength($\eta$) in region II\\
 \hline
1  & 3.53 & 24 & 0.66 &  0.570\\
0.8 & 2.97 & 27 & 0.62 & 0.500\\
0.7 & 2.05 & 30 & 0.48 &  0.450\\
0.6 & 1.78 & 34 & 0.47 &  0.420\\
0.5 & 1.71 & 42 & 0.56 &  0.395\\
0.4 & 1.65 & 45 & 0.57 &  0.340\\
0.3 & 1.41 & 50 & 0.55 &  0.300\\
0.2 & - & - & - &  0.240\\
0.15 & - & - & - & 0.210\\
0.1 & - & - & - &  0.190\\
0.05 & - & - & - & 0.170\\
0.01 & - & - & - & 0.140\\
 \hline
\end{tabular}
\caption{ Different numerical results for different distance dependent parameter ($a_{0}$). Here density within the band ($\rho_{b}$), which decreases with $a_{0}$ and width of the band ($W_{b}$), which increases as we decrease $a_{0}$. Also there is no clear band formation for $a_{0}< 0.3$ that's why we are not calculating $\rho_{b}$ and $W_{b}$ for $a_{0}<0.3$. Fraction of particles participating in band formation, that is almost same for all $a_{0}$ (47$\%$ to 66$\%$  particles are participating in band formation). We have also given $\eta$ values we have used to do all these calculation in region II of phase transition plot of global velocity in inset of Fig \ref{fig:fig7}. System parameters we have used is same as Fig \ref{fig:fig2b} .}
\label{table:table1}
\end{table*}

\begin{figure}[ht]
\centering
\includegraphics[width=1.0\linewidth]{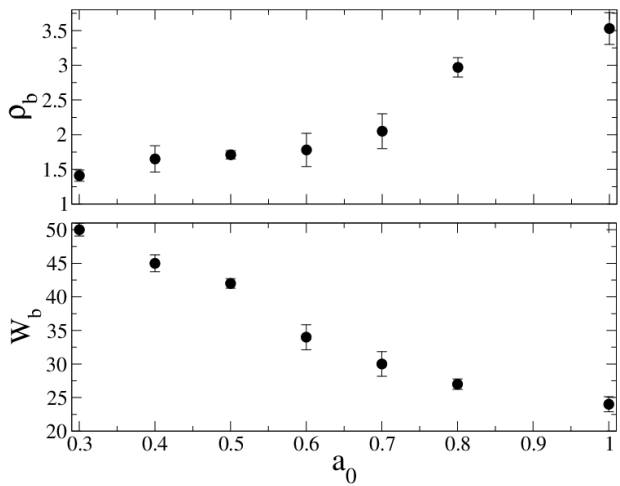}
\caption{Upper panel : Plot of band density ($\rho_{b}$) vs. distance dependent parameter ($a_{0}$). $\rho_{b}$ increases as we increase $a_{0}$. Lower panel : We plot band width ($W_{b}$) vs. $a_{0}$ and $W_{b}$ increases as we decrease $a_{0}$. For $a_{0} <0.3$ there is no clear band formation. System parameters we have used is same as Fig \ref{fig:fig2b} .}
\label{fig:fig3}
\end{figure}

\begin{figure}[ht]
\centering
\includegraphics[width=1.0\linewidth]{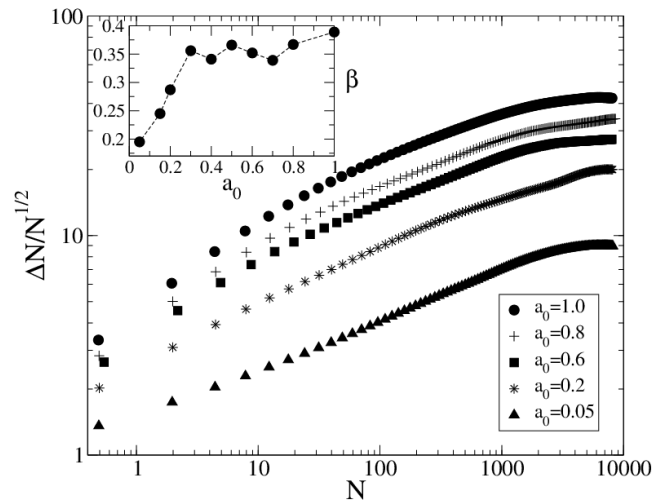}
\caption{Plot of $\Delta N/N^{\frac{1}{2}}$ vs. N for different distance dependent parameter ($a_{0}$=1.0,0.8,0.6,0.2,0.05). In the inset we plot the exponent $\beta$ vs. $a_{0}$. For $a_{0}<0.3$, $\beta$ increases as we increase $a_{0}$, for $0.3<a_{0}<0.8$ there is a plateau region and for $a_{0}> 0.8$ it increases again. For equilibrium system exponent $\beta=0.0$. System parameters we have used is same as Fig \ref{fig:fig2b} . }
\label{fig:fig5}
\end{figure}

\begin{figure}[ht]
\centering
\includegraphics[width=1.0\linewidth]{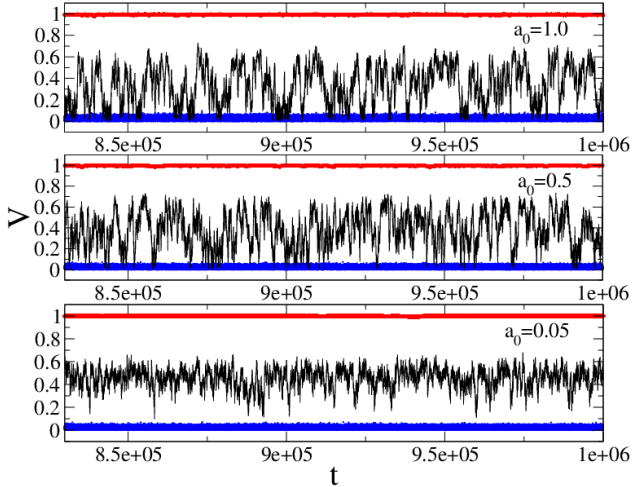}
\caption{(Color online:) Plot of global velocity (V) vs. time for three different distance dependent parameter ($a_{0}=1.0,0.5,0.05$) in time span of $2 \times 10^{5}$ from top to bottom in decrasing order. For each $a_{0}$ We choose three different region in phase transition plot of global velocity V , region I, region II and region III, are shown in inset of  Fig \ref{fig:fig7}. Here we are using different color for three different region, blue one for region I, black one for region II and  red one for region III. System parameter we have used here, $L=50, \rho_0=1.0$ . }
\label{fig:fig4}
\end{figure}

\begin{figure}[ht]
\centering
\includegraphics[width=1.0\linewidth]{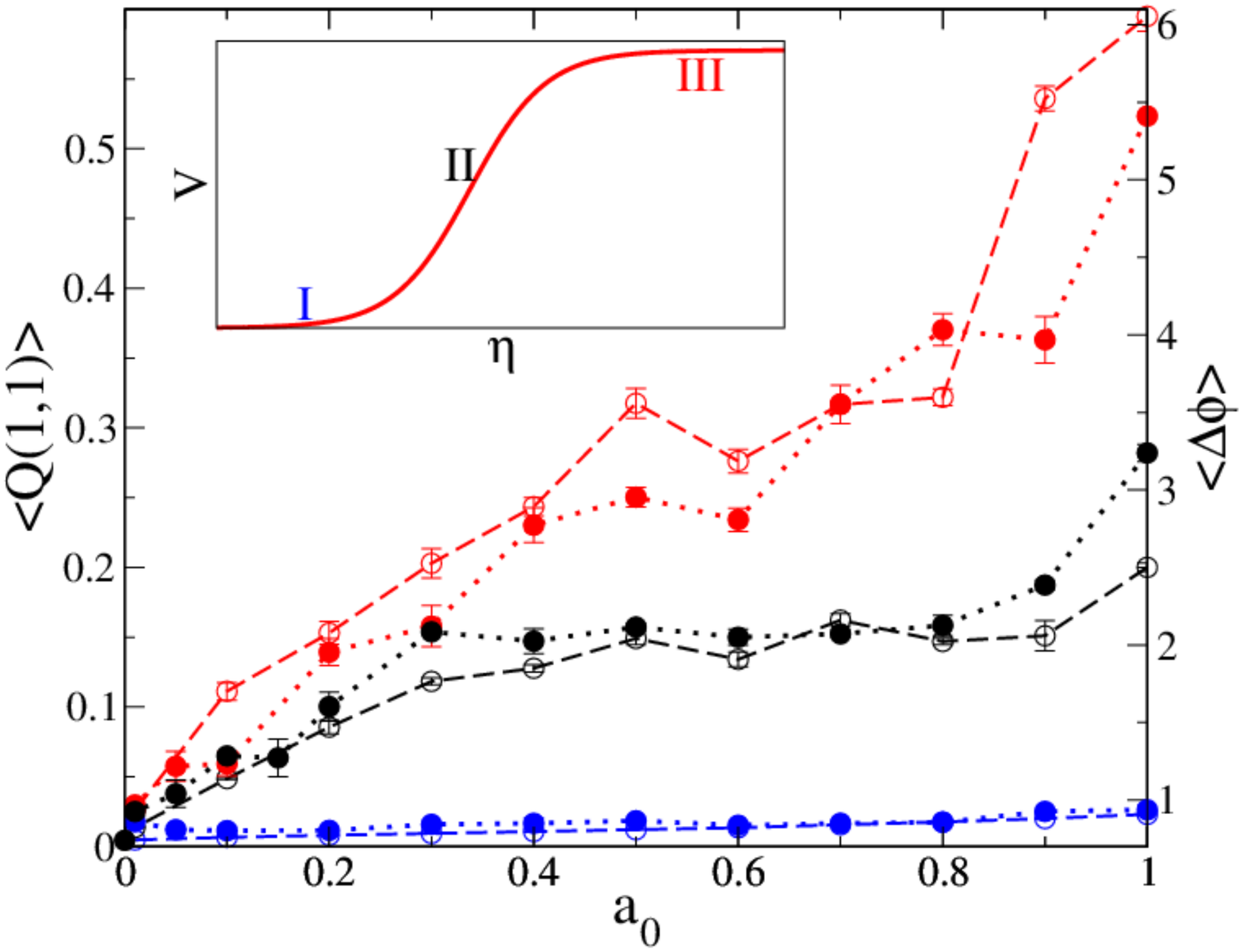}
\caption{(Color online:) Plot of average density Phase separation order parameter along diagonal direction ($<Q(1,1)>$) and $<\Delta \phi>$ (defined in text) vs. distance dependent parameter ($a_{0}$) for three different region in global velocity V phase transition plot, are shown inset of Fig \ref{fig:fig7} , blue one for region I, black one for region II and red one for region III.  In this figure left side we have labelled for $<Q(1,1)>$ and right side we have labelled for $ < \Delta \phi > $ . We have used dotted line for $<Q(1,1)>$ and dashed line for $<\Delta \phi> $ plot.  In both plot for smaller values of $a_{0}<0.3$ , $<Q(1,1)>$ and $< \Delta \phi>$ approches to zero, for $a_{0}>0.3$ and $a_{0}<0.8$ there is plateau region and for large value of $a_{0}>0.8$ to $a_{0}=1.0$ there is sharp increase in value of both $<Q(1,1)>$ and $< \Delta \phi>$. System parameters we have used here is same as Fig \ref{fig:fig2b} .}
\label{fig:fig7}
\end{figure}

\section{Results \label{results}}
We first study our model in region $II$ of phase diagram: 
Each particle is chosen one 
by one and sequencially  position and orientation 
are updated  using Eqs. \ref{eqn1} and \ref{eqn2}.
Position
and orientation of particles are stored in steady state, which we 
check by consistency of instantaneous  global order 
parameter.  Typical snapshot of particle's
position for four different values of $a_0=1.0,0.7,0.5,0.3$ at steady state and when there is clear band,
is shown in Fig \ref{fig:fig2b}  (upper pannel). 
 One of the main characteristic of polar flock, is the formation of bands
in ordered state also obtained in previous study of Chate et al. \cite{chate2007} ,\cite{chate2008}.
Similar bands are  found in other microscopic models \cite{biplab} 
as well as coarse-grained studies \cite{shradhapre}.
Our model reduces to the Vicsek's type   for $a_0=1.0$, where we also
find clear bands
as shown in Fig \ref{fig:fig2b}. 
As we vary $a_0$, size of the band
increases as shown by increasing size of horizontal bar. Mean alignment
of particles inside the band is perpendicular to the long axis of the band and bands 
typically move in one direction. Direction of motion of band is shown by
big arrow in the Fig \ref{fig:fig2b}.  In Fig \ref{fig:fig2b} (lower panel),
 we plot the one dimensional distribution of density along the band direction and average
 over other direction. For $a_0=1.0$ density distribution shows sharp peak and width is small. As we decrease
$a_0$, height of peak decreases and width increases. 
In Fig \ref{fig:fig3} we plot width of the band $W_b$, calculated from the width of the one dimensional density
distribution, averaged over many snapshots. Mean density inside the band which we  define as 
$\rho_{b} = \frac{N_{b}}{W_{b} \times L}$,
where $N_{b}$ is the number of particles participate in band formation and $W_{b}$ is the width of the band.
 As shown in Fig \ref{fig:fig3} , mean density of particles inside the band
increases as we increase $a_0$ and width of the band decreases with increasing $a_0$. In table \ref{table:table1}
we show the variation of mean width of band $W_{b}$, mean density inside the band $\rho_{b}$ and fraction of
particles participate for  band formation $n_{c}= \frac{N_{b}}{N} $ for different distance dependent parameter $a_0$. We find
although  both $W_{b}$ and $\rho_{b}$ shows variation as we change $a_0$, but  $n_{c}$ does not show any
systematic change as we decrease $a_0$. It varies from $0.66$ (66 $\%$ particles) to $0.47$ (47 $\%$ particles).
Also for $a_0 < 0.3$, there is no clear band formation, hence it is not possible to calculate different quantities (e.g $W_b$, $\rho_b$, etc.). 
Hence as we decrease $a_0$ system shows a change from high density narrow bands to low density wide bands and
finally for very small $a_0$, there is no band.\\
Formation of high and low density bands should also be visible in two-point density structure factor.
Finite size of bands or clusters show a presence of critical wavevector in the system. We calculate the
two-point density structure factor $S({\bf q})$ using 
linearised calculation in ordered state. Details of calculation are given in appendix \ref{App:AppendixA}.
$S({\bf q})$ is calculated in the direction of ordering or
along the bands direction. From Eqs.\ref{eqn25} we 
 find $S({\bf q})= \frac{v_{0}^{2}\rho_{0}^{2} \triangle_{0}}{c_{2}}[\frac{1}{q^{2}+q_{1}^{2}}+\frac{1}{q^{2}-q_{2}^{2}}] $
 In the above expression of $S({\bf q})$, $c_{2}$ is a constant, is defined in
\ref{eqn13} and expression for $q_1$ and $q_2$ is given in Eqs.\ref{eqn28} and Eqs.\ref{eqn29} respectively.  
$S({\bf q})$ diverges at critical wavevector 
$ q_c =q_{2}= \sqrt{\frac{C B_{+} \alpha_{1}^{\prime}}{D_{V}v_{0}^{2}v_{1}^{2}}}$.
 Where $C=v_{0}V_{0}$ and $B_{+}$  is  defined in Eqs. \ref{eqn30}.
Critical length scale
 $L_c = q_c^{-1} \simeq \sqrt{\frac{D_{V}v_{0}^{2}v_{1}^{2}}{C B_{+} \alpha_{1}^{\prime}}}$
 decreases  with increasing $\alpha_1'=\frac{d\alpha}{d\rho}\vert_{\rho_0}$, which  depends on
the density dependence of $\alpha(\rho)$. For $\alpha_1'=0$ or $\alpha$ is independent of density and 
 Eqs. \ref{eqn6} reduces to Toner and Tu \cite{tonertu}.
Variation of width of the band $W_b$ in microscopic  simulation with distance dependent 
parameter $a_0$ and dependence of critical wavevector $q_c$ with  $\alpha_1'$, shows one to one 
mapping between them.\\
Band formation in polar flock  or clustering
of particles also implies the  density phase separation.
In order to calculate density phase separation we first calculate fluctuation 
in density in cells of small size.  To calculate such quantity,  whole  system is divided 
into  $N_c$ small cells of size $1 \times 1$. Hence for  system of size $L \times L$, there will 
be $L^{2}$ small cells ($N_{c}=L^{2}$). We calculate number of particles in each cell 
in the steady state. Then  standard deviation in particle number is calculated
from different cells, 
which is defined by $\Delta \phi$ 
\begin{equation}
\Delta \phi =\sqrt{\frac{1}{N_{c}}\sum_{j=1}^{N_{c}}(\phi_{j})^{2}-(\frac{1}{N_{c}}\sum_{j=1}^{N_{c}}\phi_{j})^{2}}
\label{eqn27}
\end{equation}
where, $\phi_{j}$ is the number of particles in $j$th cell.

We also calculate Fourier transform of density defined by
\begin{equation}
Q({\bf k})=\mid \frac{1}{L} \sum _{i,j=1} ^{L} e^{i{\bf k}\cdot {\bf r}} \rho(i,j) \mid
\label{eqn28}
\end{equation}
where ${\bf k}=\frac{2 \pi (m, n)}{L}$, where $m,n$=$0$, $1$, $2$ ...., $L-1$ are a two dimensional wave vector. 
We choose directions $(1,0)$ and $(0,1)$
as two directions of square box hence direction $(1,1)$
is along the diagonal of square box.
During evolution
of flock, direction of band changes with time. Hence to get maximum information
about the clustering we calculate first non-zero value of $Q({\bf k})$ 
 in the diagonal direction or
$Q(1,1)$, where  $m=n=1$. Both  $\Delta \phi(t)$ and $Q(1,1)(t)$ are
calculated at different times in the steady state. Then we average it over large time and
calculate $<\Delta \phi>$ and $<Q(1,1)>$. Plot of $<\Delta \phi>$ and 
$<Q(1,1)>$ vs. $a_0$ is  shown on right and 
left respectively of Fig. \ref{fig:fig7}. $<\Delta \phi>$ and $<Q(1,1)>$ is calculated in all
three  regions of phase diagram  (inset of Fig. \ref{fig:fig7}). 
For region I in the phase diagram, 
 where system is in the disordered state, both 
$<\Delta \phi>$ and  $<Q(1,1)>$ remains small hence no phase separation. 
For region II or close to order-disorder transition, as we increase $a_0$, first
both $\Delta \phi$ and  $<Q(1,1)>$ increases with $a_0$, then
shows a plateau type behaviour for $0.3< a_0 <0.7$ and again 
increases for $a_0 > 0.7$. Hence system shows  no phase separation 
for small $a_0$ and then gradually goes to moderate phase separation and finally for large
$a_0$ shows strong  phase separation.
We find similar results for region III of phase diagram, where system is in
the deep ordered state.
Hence density  shows another phase  transition from phase separated to nonphase separated
state  as a function of distance dependent parameter $a_0$.\\

\begin{figure*}[ht]
\centering
\begin{minipage}{.5\textwidth}
  \includegraphics*[width=1.0\linewidth]{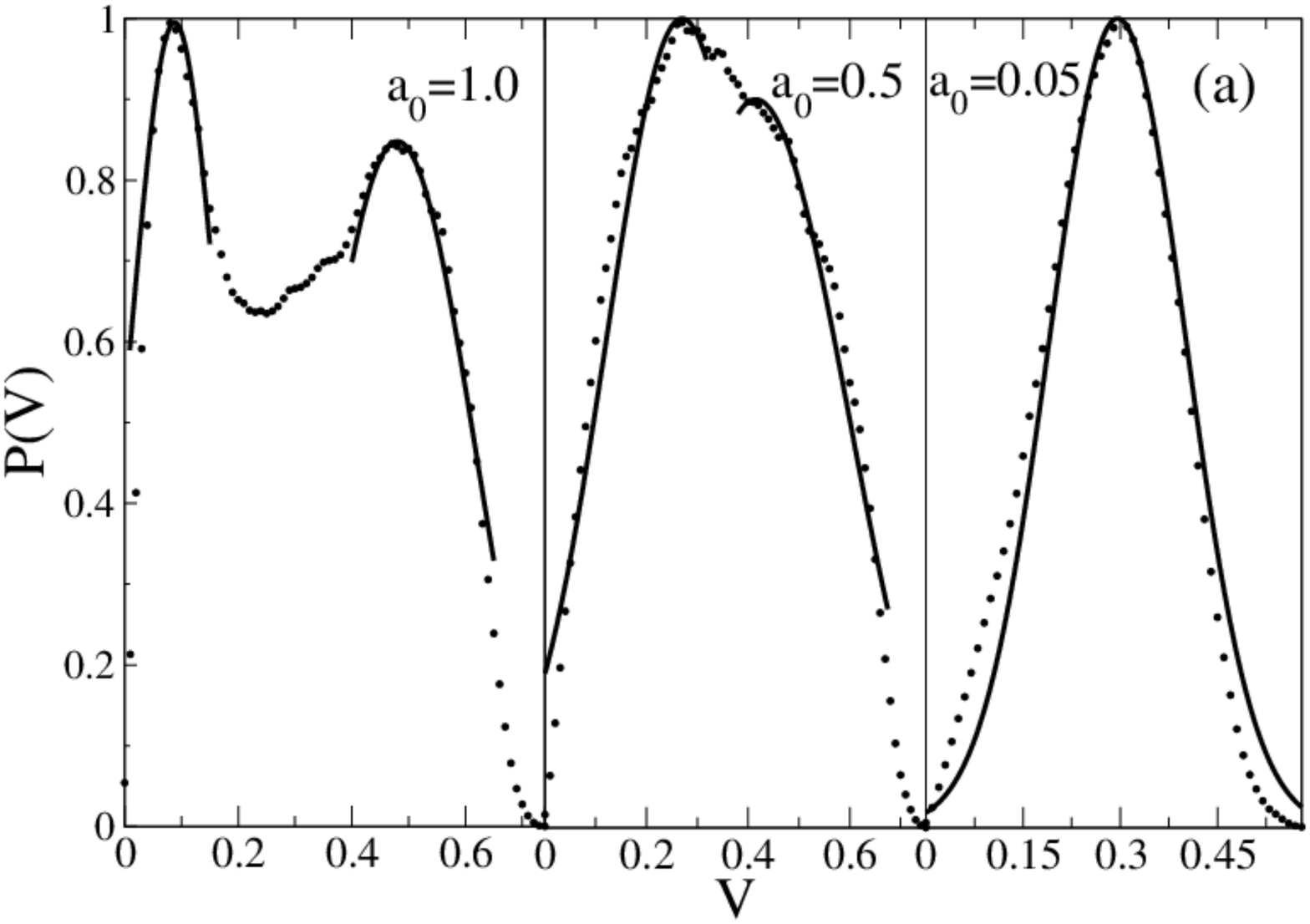}
\end{minipage}%
\begin{minipage}{.5\textwidth}
  \includegraphics*[width=1.0\linewidth]{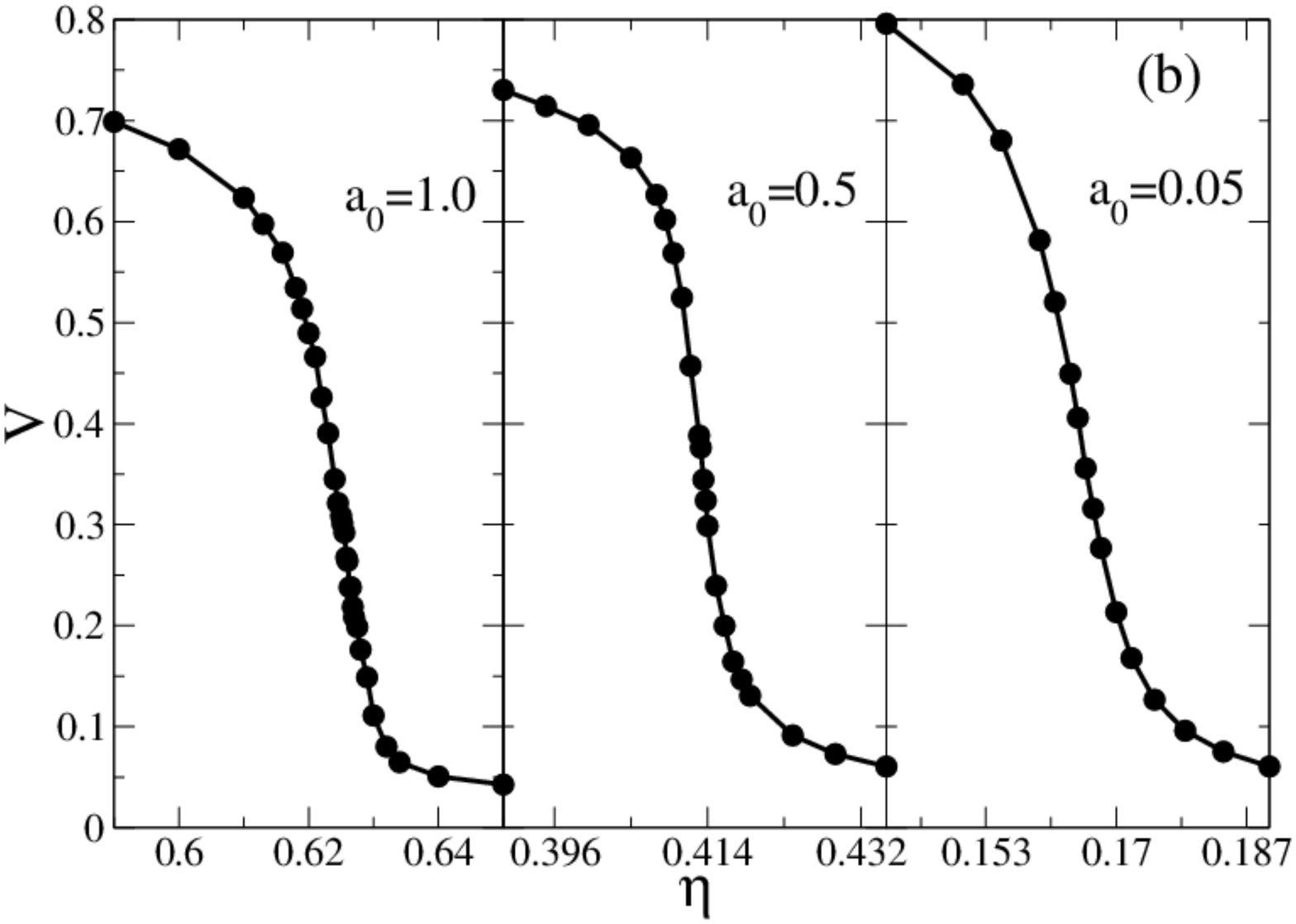}
\end{minipage}
\begin{minipage}{.5\textwidth}
  \includegraphics*[width=1.0\linewidth]{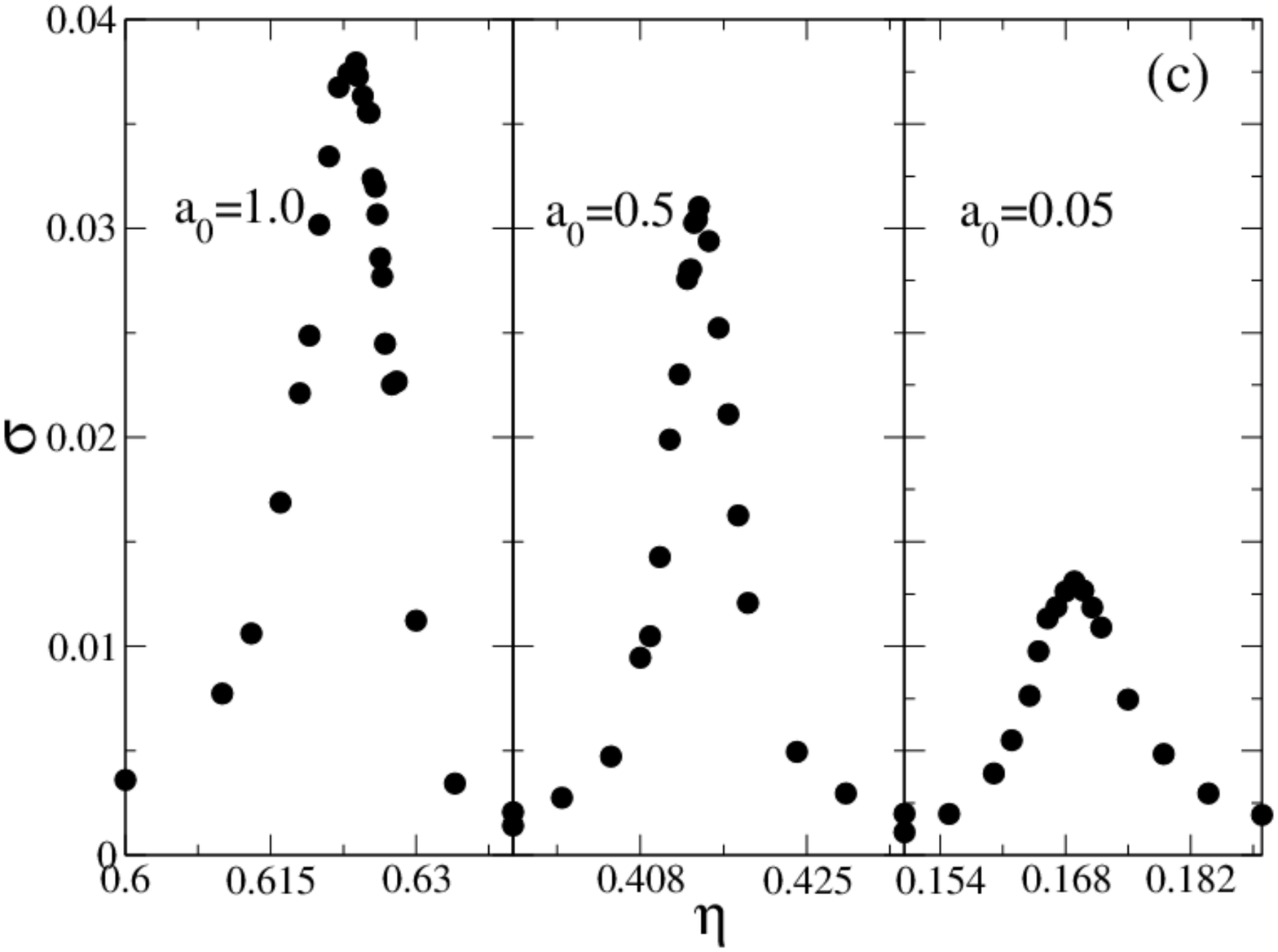}
\end{minipage}%
\begin{minipage}{.5\textwidth}
  \includegraphics*[width=1.0\linewidth]{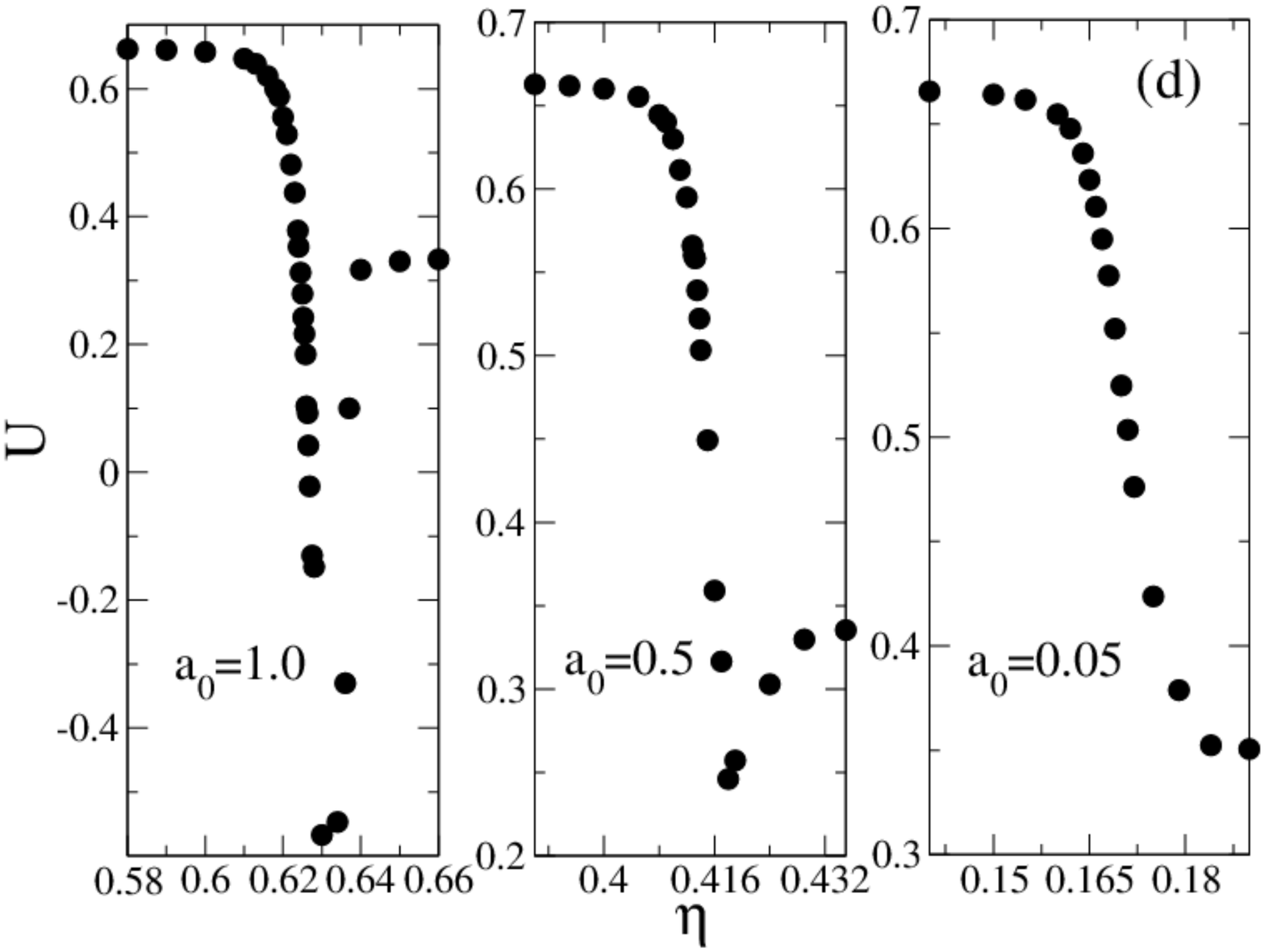}
\end{minipage}
\caption{Upper panel: In figure (a) we plot probability distribution of Global velocity V for $a_{0}=1.0,0.5,0.05$ from left to right in decreasing order. Distribution of V changes bimodal to unimodal as we decrease $a_{0}$. In figure (b) we plot phase transition curve of V vs. noise strength $\eta$ in same order as in figure (a). Change in V becomes more and more sharp as we increase $a_{0}$. Lower panel: We plot variance ($\sigma$) of V vs. $\eta$ for $a_{0}=1.0,0.5,0.05$ from left to right in figure (c). $\sigma$ decays both side of critical noise strength $\eta_{c}$ for all $a_{0}$. In figure (d) we plot Binder cumulant (U) vs. $\eta$ in same order as in figure (c). For $a_{0}=1.0$ there is sharp jump in variation of U from ordered state ($U=2/3$) to disordered state ($U=1/3$), for $a_{0}=0.5$ jump in variation of U becomes less sharp and for $a_{0}=0.05$ variation of U becomes continuous from $U=2/3$ to $U=1/3$. System parameters we have used is same as Fig \ref{fig:fig4} .}
\label{fig:fig6}
\end{figure*}

We also calculate density fluctuation $\Delta N = \sqrt{<N^2> - <N>^2}$, as we vary distance
dependent parameter $a_0$. We find for all $a_0$, density fluctuation $\Delta N/N^{1/2} \simeq N^{\beta}$
and $\beta >0.0$, hence density fluctuation is larger than thermal  equilibrium system, where $\beta=0.0$.
 Large density fluctuation is one of the characteristic feature of active self-propelled systems \cite{chate2004, shradhaprl, das2012, aditi}.
But we find the exponent $\beta$ varies as we tune $a_0$.  In the inset of Fig \ref{fig:fig5}, we plot
variation of exponent $\beta$ for different $a_0$. $\beta$ vs. $a_0$ plot
is almost flat with $\beta \simeq 0.4$ for $a_0 > 0.3$ and approaches zero for very small $a_0$. 

{\it Phase Transition :}
Now we characterise the order-disorder transition for three different values of
distance dependent parameter $a_0 = 1.0$, $0.5$ and $0.05$. As shown in Fig \ref{fig:fig7}
these three values of $a_0$ are three different points in density phase separation curve. For $a_0=1.0$,
system is strongly phase separated, for $a_0=0.5$, moderate phase separation (plateau region) and for
$a_0=0.05$, no phase separation.  In Fig \ref{fig:fig4}  we first plot the time series of global
velocity for three different regions, region I, region II and region III as shown in inset of Fig \ref{fig:fig7}.
During a large span of time ($t=2 \times 10^5$) in the steady state (after $t =  8 \times 10^5$), 
global velocity $V$ approches to 
value close to $1$ for all three $a_{0}$ in region $III$ and for 
region $I$ it approaches to $0$. But  for region II system shows
switching type behaviour, where system continuously switches from ordered $V \simeq 0.6$ to
disordered $V \simeq 0.1$ state for $a_{0}=1$. Such switching behaviour of global velocity is also 
observed in previous study of  \cite{chate2008}  ,\cite{biplab} ,\cite{shradhapre} 
As we tune $a_0=0.5$, time series of global velocity shows weaker switching behaviour and for very small
$a_0=0.05$ it shows huge fluctuation with no switching (shown in Fig \ref{fig:fig4}). 
 Probability 
distribution of global velocity $P({V})$ shows bistable behaviour for $a_0=1.0$ and gradually
switches to unimodal behaviour for small $a_0$ as shown in Fig \ref{fig:fig6} (a). We have shown phase transition curve of V for 
three different $a_{0}$ in Fig \ref{fig:fig6} (b). For larger value of $a_{0}$ change in V with noise strength $\eta$ becomes very sharp and as we 
decrease $a_{0}$, change in V with $\eta$ becomes more and more continuous. We also calculate
variance of order parameter $\sigma=\langle V^{2} \rangle-\langle V \rangle^{2}$, shown in Fig \ref{fig:fig6} (c) and also
the fourth order binder cumulant defined by $U=1-\frac{<V^{4}>}{3<V^{2}>^{2}}$, Fig \ref{fig:fig6} (d)
 shows strong discontinuity from $1/3$ for disordered state to $2/3$ for ordered state
as we approach critical $\eta$ for $a_0=1.0$ and discontinuity decreases with $a_0$ and smoothly 
goes from disordered value $1/3$ to ordered state value $2/3$ for $a_0=0.05$. System
shows a transition from disordered to ordered state for all $a_0$, but nature of transition changes from first
order type to continuous as we tune $a_0$.  Also density changes from phase separated 
to nonphase separated state.

\section{Discussion \label{discussion}}
We studied a collection of polar self-propelled  particles, interacting
 through distance dependent short range alignment interaction. Such distance
dependent model is biologically motivated, where particles interact more strongly with
their closest neighbours. Distance dependent interaction is introduced
through an interaction parameter $a_0$, which varies from $1.0$ to $0.0$. For $a_0=1.0$, model
reduces to Vicsek's type  and $a_0=0.0$, implies no interaction. For all $a_0$'s
system shows a phase transition from disordered (global velocity $V=0.0$) to
ordered (finite global velocity) as we vary noise intensity $\eta$. For large $a_0$ density
shows formation of bands, characteristic of bands changes  with $a_0$. For $a_0 $ close to
$1.0$, bands are strong with small width and high density and as we decrease $a_0$, bands 
become weak. 
Our numerical
result is consistent with analytical calculation of density structure factor as
shown in Eqs. \ref{eqn25}. Where we find a critical wavevector at which structure factor diverges.
The  critical wave vector decreases or wavelength increases with decreasing $\alpha_{1}^{\prime}$.\\
Our finding in the work shows that density phase separation plays an important role in
determining the nature of phase transition. First order phase transition in
Vicsek's model is because of strong density phase separation. 
Density phase separation in microscopic simulation can be tune in many ways,
we can use distance dependent interaction, which controls the number of interacting
neighbours. Changing the speed of particle in the Vicsek's model will give the same result.
For large speed system should show phase separated state and first order transition and for 
small speed we will have nonphase separated or continuous transition. 
Microscopic model we introduce here is not a unique distance dependent model,
other models where interaction vary with other functional dependence on distance
will also show the similar results.\\    
\begin{acknowledgments}
{S. Pattanayak would like to thank Dr. Manoranjan Kumar for his 
kind cooperation and useful suggestions through out this work.
S. Pattanayak would like to thank Department of Physics, IIT (BHU), Varanasi for
 kind hospitality. 
S. Pattanayak would also like to thank Mr. Rakesh Das for his useful suggestions. 
S. Mishra would like to support DST for their partial financial support
in this work.}
\end{acknowledgments} 

\begin{widetext}
\appendix
\section{Linearised study of the broken symmetry state} \label{App:AppendixA}

The hydrodynamic equations Eqs.\ref{eqn5} and \ref{eqn6}, admit two homogeneous solutions: an isotropic state with ${\bf V}=0$ for $\rho < \rho_{c}$ and a homogeneous polarized state with ${\bf V}=V_{0}{\bf x}$ for $\rho > \rho_{c}$, where ${\bf x}$ is the direction of ordering. 
.We are mainly interested in the symmetry broken phase, specifically along the horizontal direction, in which direction large cluster or bands are moving. For $\alpha(\rho) > 0$, We can write the polarization field as, ${\bf{V}}=(V_{o}+\delta V_{x}){\bf x}+\delta {\bf V}_{y}$, where ${\bf x }$ is the direction of band formation or horizontal direction and ${\bf y}$ is the perpendicular direction, $V_{0}{\bf x}=<{\bf{V}}>$ is the spontaneous average value of $\bf{V}$ in ordered phase. We choose $V_0=\sqrt{\frac{\alpha(\rho_{0})}{\beta}}$ and $\rho = \rho _{0}+\delta \rho$ where $\rho _{0}$, coarse-grained density. Combining the fluctuations we can write in a vector format,
\begin{equation}
\delta X_{\alpha}({\bf r},t)=\left[ \begin{array}{c} \delta \rho \\ \delta V _{x} \\ \delta V_{y} \end{array} \right]
\label{eqn8}
\end{equation}

Now we introduce fluctuation in hydrodynamic equation for density then Eqs. \ref{eqn5} will reduce to,
\begin{equation}
\partial_{t} \delta \rho + v_{0}V_{0}\partial_{x} \delta \rho +v_{0}V_{0}\partial_{y} \delta \rho +v_{0}\rho_{0}\partial_{x} \delta _{V_{x}}+v_{0}\rho_{0}\partial_{y} \delta _{V_{y}}=0
\label{eqn10}
\end{equation} 
Similarly  we introduce fluctuation in velocity Eqs. \ref{eqn6}, we are writing velocity fluctuation equations for horizontal direction or direction of band formation and for perpendicular direction. We are writing fluctuation equations for both direction separately, here x-is the direction of ordering and y is perpendicular direction. We have done Taylor series expansion of $\alpha(\rho)$  in Eqs.\ref{eqn6} at $\rho=\rho_0$ and we have consider upto first order derivative term of $\alpha(\rho)$.
\begin{equation}
\partial _{t} {\delta {V _{x}}}= (\alpha (\rho _{0})+ \alpha {_{1}^{\prime}} ({\rho _{0}})\delta \rho) (V_{0} + \delta V_{x})  -\beta (V_{0}^{2} +2V_{0} \delta V_{x})(V_{0}+ \delta V_{x})  -\frac{v _{1}}{2\rho _{0}} \partial _{x}  \delta \rho +D_{V} \partial _{x} ^{2} \delta V _{x}+D_{V} \partial _{y} ^{2} \delta V _{x} -\lambda V _{0} \partial _{x} \delta V_{x} + {f _{Vx}}
\label{eqn11}
\end{equation}
\begin{equation}
\partial _{t} {\delta {V _{y}}}=(\alpha (\rho _{0})+ \alpha {_{1}^{\prime}} ({\rho _{0}})\delta \rho) (\delta V_{y})-\beta (V_{0}^{2} +2V_{0} \delta V_{x})(\delta V_{y})-\frac{v _{1}}{2\rho _{0}} \partial _{y} \delta \rho +D_{V} \partial _{x} ^{2} \delta V _{y} +D_{V} \partial _{y} ^{2} \delta V _{y}-\lambda V _{0} \partial _{x} \delta V_{y} + {f _{Vy}}
\label{eqn12}
\end{equation}
where $\alpha_{1}^{\prime}=\frac{\partial \alpha}{\partial \rho}\mid _{\rho _{0}}$ also $\lambda$ is combination of three $\lambda's(\lambda=\lambda_{1}+\lambda_{2}+2 \lambda_{3})$ terms. \\
 Now we are introducing Fourier component, $\delta  \hat{\bf X}_{\alpha}({\bf k},t)=\int exp^{i({\bf k.r}-\omega t)} \delta X_{\alpha}({\bf r},t)$ in above three fluctuation equations \ref{eqn10}, \ref{eqn11} and \ref{eqn12}. Then the coupled equations we write in matrix form.
\begin{multline}
\left[ \begin{array}{ccc} 
-i\omega+iv_{0} V_{0}q_{x}+iv_{0} V_{0}q_{y} & iv_{0}\rho_{0}q_{x} & iv_{0}\rho_{0}q_{y} \\ i \frac{v _{1}}{2 \rho _{0}} q_{x} -\alpha {_{1}^{\prime}}(\rho _{0}) V _{0} & -i \omega  +2\alpha(\rho_{0}) +D _{V}\vert q^{2}\vert +i\lambda V _{0}q _{x} & 0 \\ i \frac{v _{1}}{2 \rho _{0}} q_{y} & 0 & -i \omega  +2\alpha(\rho_{0}) +D _{V}\vert q^{2}\vert +i\lambda V _{0}q _{x}\end{array} \right]  \times \left[ \begin{array}{c} \delta \rho \\ \delta V _{x} \\ \delta V_{y} \end{array} \right] =\left[ \begin{array}{c} 0 \\ f _{Vx} \\ f_{Vy}\end{array} \right]
\label{eqn16}
\end{multline}
Earlier study \cite{shradhapre} finds horizontal fluctuation or fluctuation in the direction of band formation is important when system is close to transition. Here in our numerical study we take region II, near to transition point, shown in the inset of Fig. \ref{fig:fig7}. So, we are only considering fluctuation in ordering direction, then the above 3x3 matrix \ref{eqn16} reduce to,
\begin{equation}
\left[ \begin{array}{cc}  -i\omega+iv_{0}V_{0}q & iv_{0}\rho_{0}q \\ i \frac{v _{1}}{2 \rho _{0}} q -\alpha {_{1}^{\prime}}(\rho _{0}) V _{0} & -i \omega  +2\alpha +D _{V}q^{2} +i\lambda V _{0}q  \end{array} \right]  \times \left[ \begin{array}{c} \delta \rho \\ \delta V _{x} \end{array} \right] = \left[ \begin{array}{c} 0 \\ f _{Vx} \end{array} \right] 
\label{eqn17}
\end{equation}

We first determine the eigen frequencies of $\omega(\bf q)$ of these coupled equations and we find,
\begin{equation}
\omega _{\pm}=c_{\pm} q-i\varepsilon_{ \pm}
\label{eqn19}
\end{equation} \\
where, the sound speeds, 
\begin{equation}
c_{\pm}=\frac{1}{2}(\lambda+v_{0})V_{0}\pm c_{2}
\label{eqn18}
\end{equation}
with
\begin{equation}
c_{2}=\sqrt{(\frac{1}{4}(\lambda -v_{0})^{2}V_{0}^{2}+v_{0}v_{1})}
\label{eqn13}
\end{equation}
and the damping $\varepsilon _{\pm}$ in the Eqs. \ref{eqn19} are $O({\bf q^{2}})$  and given by, 
\begin{equation}
\varepsilon _{\pm}=\pm \frac{c_{\pm}}{2c_{2}}[2\alpha+D_{V}q^{2}]\mp\frac{1}{2c_{2}}[2\alpha v_{0}V_{0}+v_{0}V_{0}\alpha _{1}^{\prime}+v_{0}V_{0}D_{V}q^{2}]
\label{eqn14}
\end{equation}
Here, important thing is that unlike isotropic problem $d>2$ there are no transverse mode, we always have just two longitudinal goldstone modes, associated with $\delta \rho$ and $V_{x}$.

Now the two-point density auto correlation along the direction of ordering,
\begin{equation}
\centering
C_{\rho \rho} ({\bf q},\omega)=\frac{v_{0}^{2}\rho _{0}^{2}2 \triangle _{0} q^{2}}{(-\omega ^{2}+(v _{0}+\lambda)V _{0}q\omega-v _{0}\lambda V_{0}^{2}q^{2}+\frac{v_{1}v_{0}q^{2}}{2} )^{2}+[\omega (2\alpha + D_{V}q^{2})-  q(2\alpha v_{0}V_{0}+v_{0}V_{0}\alpha _{1}^{\prime}+v_{0}V_{0}D_{V}q^{2})]^{2}}
\label{eqn20}
\end{equation} 
\begin{equation}
\centering
C_{\rho \rho} ({\bf q},\omega)=\frac{v_{0}^{2}\rho _{0}^{2} 2\triangle _{0}q^{2}}{(\omega -c_{+}q)^{2}(\omega -c_{-}q)^{2}+[\omega (2\alpha + D_{V}q^{2})-q(2\alpha v_{0}V_{0}+v_{0}V_{0}\alpha _{1}^{\prime}+v_{0}V_{0}D_{V}q^{2})]^{2}}
\label{eqn32}
\end{equation}
If we plot density-density correlation function as a function of $\omega$ there are two peaks at $\omega=c_{\pm}q$.
From above density auto correlation it is very straight forward to calculate structure factor.
\begin{equation}
S(q,t)=\frac{1}{2\pi}\int _{-\infty} ^{+\infty}\left\langle \vert \delta \rho (q,\omega ) \vert ^{2} \right\rangle d\omega=\frac{1}{2\pi}\int _{-\infty} ^{+\infty} C_{\rho \rho} ({\bf q},\omega) d\omega
\label{eqn21}
\end{equation} 
\begin{equation}
S(q,t)=\frac{1}{2\pi}\int _{-\infty}^{+\infty}\frac{v_{0}^{2}\rho _{0}^{2} 2\triangle _{0}q^{2}}{(\omega -c_{+}q)^{2}(\omega -c_{-}q)^{2}+[\omega (2\alpha + D_{V}q^{2})-q(2\alpha v_{0}V_{0}+v_{0}V_{0}\alpha _{1}^{\prime}+v_{0}V_{0}D_{V}q^{2})]^{2}}
\label{eqn22}
\end{equation}
\begin{equation}
S(q,t)=\frac{v_{0}^{2}\rho_{0}^{2} 2 \triangle _{0} q^{2}}{2c_{2}q}[\frac{1}{c_{+}q(2\alpha+D_{V}q^{2})-q(v_{0}V_{0}(2\alpha+D_{V}q^{2})+v_{0}V_{0}\rho_{0}\alpha_{1}^{\prime})}]
\label{eqn23}
\end{equation}

\begin{equation}
S(q,t)=\frac{v_{0}^{2}\rho_{0}^{2} \triangle_{0}}{c_{2}}[\frac{1}{q^{2}+q_{1}^{2}}+\frac{1}{q^{2}-q_{2}^{2}}]
\label{eqn25}
\end{equation}
where,
\begin{equation}
q_{1}^{2}=[\frac{v_{0}V_{0}\alpha_{1}^{\prime} B_{-}}{D_{V}v_{0}^{2}v_{1}^{2}}]
\label{eqn28}
\end{equation}
\begin{equation} 
q_{2}^{2}=[\frac{v_{0}V_{0}\alpha_{1}^{\prime} B_{+}}{D_{V}v_{0}^{2}v_{1}^{2}}]
\label{eqn29}
\end{equation}
and
\begin{equation}
B_{\pm}=[\sqrt{\frac{1}{4}(\lambda -v_{0})^{2}V_{0}^{2}+v_{0}v_{1}} \mp \frac{1}{2}(\lambda - v_{0})V_{0}]
\label{eqn30}
\end{equation}
 All the constants in wave vector expression is defined earlier. From there it is very clear $q_{1}^{2}$ and $q_{2}^{2}$ are positive. Now from above structure factor expression we get critical wave vector below which structure factor diverges, 
\begin{equation}
D_{V}q_{2}^{2}=[\frac{v_{0}V_{0}\rho_{0}\alpha_{1}^{\prime} B_{+}}{v_{0}^{2}v_{1}^{2}}]-2\alpha B_{+}
\label{eqn26}
\end{equation}
When the system is near to critical point then $\alpha(\rho_{0})$ will be close to zero. Then we can write expression for critical wave vector,
\begin{equation}
 q_{c}=q_2=\sqrt{\frac{C B_{+} \alpha_{1}^{\prime}}{D_{V}v_{0}^{2}v_{1}^{2}}}
\label{eqn27}
\end{equation}
Where,
$C=v_{0}V_{0}$ 
Here for $\alpha _{1}^{\prime} \neq 0  $ we get a critical wave vector $q_{c}$ at which structure factor diverges and it also gives a critical length scale $L_{c}$ of the system. Where, $\alpha_1'$ is density dependent alignment term, which is similar to the distance dependence parameter $a_{0}$ in our numerical study and for  $\alpha _{1}^{\prime} = 0  $ our study reduces to Toner and Tu study \cite{tonertu}. 
\end{widetext}




\end{document}